\documentclass{article} 
\usepackage{iclr2025_conference,times}

\usepackage{amsmath,amsfonts,bm}

\def\eqref#1{equation~\ref{#1}}

\def\1{\bm{1}}

\DeclareMathAlphabet{\mathsfit}{\encodingdefault}{\sfdefault}{m}{sl}
\SetMathAlphabet{\mathsfit}{bold}{\encodingdefault}{\sfdefault}{bx}{n}

\usepackage{graphicx} 
\usepackage{times}

\usepackage{hyperref}
\usepackage{url}
\usepackage{bbm}
\usepackage{graphicx}

\usepackage{amsthm}
\theoremstyle{plain}

\usepackage{comment}
\usepackage[utf8]{inputenc} 
\usepackage[T1]{fontenc}    
\usepackage{hyperref}       
\usepackage{url}            
\usepackage{booktabs}       
\usepackage{amsfonts}       
\usepackage{nicefrac}       
\usepackage{microtype}      
\usepackage{xcolor}         
\usepackage{multirow} 
\usepackage{graphicx}
\usepackage{booktabs}
\usepackage{tabularx}
\usepackage{subcaption}
\usepackage{wrapfig}
\usepackage{enumitem}

\usepackage{amsmath}
\usepackage{amssymb}
\usepackage{mathtools}
\usepackage{amsthm}

\usepackage{multirow}
\usepackage{import}
\usepackage{pifont}
\usepackage{textcomp}
\usepackage{color, colortbl}
\definecolor{greyC}{RGB}{180,180,180}
\definecolor{greyL}{RGB}{235,235,235}
\usepackage{footmisc}
\usepackage[noend]{algpseudocode}
\usepackage{bm}
\usepackage[flushleft]{threeparttable}
\usepackage[nospace]{cite}
\usepackage[]{hyperref} 
\usepackage{makecell}

\usepackage[utf8]{inputenc} 
\usepackage[T1]{fontenc}    
\usepackage{hyperref}       
\usepackage{url}            
\usepackage{booktabs}       
\usepackage{amsfonts}       
\usepackage{nicefrac}       
\usepackage{microtype}      
\usepackage{xcolor}         

\usepackage{hyperref}
\usepackage{algorithm}
\usepackage{algcompatible}
\usepackage{wrapfig}

\usepackage{varwidth}
\usepackage{tikz}
\usetikzlibrary{calc}
\usetikzlibrary{positioning, shapes, fit, backgrounds, decorations.pathreplacing}

\usepackage{changepage}
\usepackage{sidecap}

\definecolor{mblue}{RGB}{0, 61, 124}
\definecolor{myellow}{RGB}{239, 124, 0}
\definecolor{mnavy}{RGB}{0,0,128}
\definecolor{minc}{RGB}{0,128,0}
\definecolor{mdec}{RGB}{255,0,0}
\definecolor{mhold}{RGB}{128,128,128}
\newcommand{\stitle}[1]{\vspace{1mm} \noindent {\bf #1}}

\newcommand{\ms}[2]{{#1\tiny{$\pm$#2}}}
\usepackage{pifont}

\title{Exploring LLM Cryptocurrency Trading Through Fact-Subjectivity Aware Reasoning \\}
\author{
Qian Wang\textsuperscript{1}  
Yuchen Gao\textsuperscript{1}
Zhenheng Tang\textsuperscript{2}
Bingqiao Luo\textsuperscript{1} 
Nuo Chen\textsuperscript{1}
Bingsheng He\textsuperscript{1} 
\vspace{0.5em} \\
\textsuperscript{1}National University of Singapore \quad 
\textsuperscript{2}Hong Kong University of Science and Technology
}

\iclrfinalcopy 
\begin{document}

\maketitle

\begin{abstract}
While many studies show that more advanced LLMs excel in tasks such as mathematics and coding, we observe that in cryptocurrency trading, stronger LLMs sometimes underperform compared to weaker ones. To investigate this counterintuitive phenomenon, we examine how LLMs reason when making trading decisions. Our findings reveal that (1) stronger LLMs show a preference for factual information over subjectivity; (2) separating the reasoning process into factual and subjective components leads to higher profits. Building on these insights, we propose a multi-agent framework, FS-ReasoningAgent, which enables LLMs to recognize and learn from both factual and subjective reasoning. Extensive experiments demonstrate that this fine-grained reasoning approach enhances LLM trading performance in cryptocurrency markets, yielding profit improvements of 7\% in BTC, 2\% in ETH, and 10\% in SOL. Additionally, an ablation study reveals that relying on subjective news generates higher returns in bull markets, while focusing on factual information yields better results in bear markets. Code is available at \url{https://github.com/Persdre/FS-ReasoningAgent}.
\end{abstract}

\section{Introduction}
Large Language Models (LLMs) demonstrate excellent reasoning abilities \citep{chang2024survey} and achieve outstanding performance in fields that require high-level reasoning, such as coding and mathematics \citep{guo2023evaluating}. Recent research also highlights their ability to interpret financial time series and improve cross-sequence reasoning \citep{wei2022chain, yu2023temporal, zhang2023multimodal, zhao2023survey, yang2024harnessing}. Furthermore, the development of LLM-based trading strategies such as Sociodojo \citep{cheng2024sociodojo} and CryptoTrade \citep{li2024reflective} highlights the exceptional reasoning capabilities of LLMs in making high-return trading decisions driven by market news.

\begin{wrapfigure}{R}{0.5\textwidth}
\begin{minipage}{0.5\textwidth}
\begin{table}[H]
    \small 
    \caption{Performance comparison of single LLMs, and baseline trading strategies on ETH during both Bull and Bear market conditions.}
    \vspace{-0.2cm}
    \resizebox{\columnwidth}{!}{ 
    \begin{tabular}{l|cc|cc|cc}
        \toprule
        \textbf{Strategy} & \multicolumn{2}{c|}{\textbf{Total Return (\%)}} & \multicolumn{2}{c|}{\textbf{Daily Return (\%)}} & \multicolumn{2}{c}{\textbf{Sharpe Ratio}} \\
        \cmidrule(lr){2-3} \cmidrule(lr){4-5} \cmidrule(lr){6-7}
        & \textbf{Bull} & \textbf{Bear} & \textbf{Bull} & \textbf{Bear} & \textbf{Bull} & \textbf{Bear} \\
        \midrule
        Buy and Hold  & 22.59 & -12.24 & \ms{0.36}{2.62} & \ms{-0.17}{2.39} & 0.14 & -0.07 \\
        SMA           & 10.17 & -10.12 & \ms{0.18}{2.29} & \ms{-0.15}{1.64} & 0.08 & -0.09 \\
        SLMA          & 5.20  & -15.90 & \ms{0.11}{2.37} & \ms{-0.24}{1.86} & 0.05 & -0.13 \\
        MACD          & 7.72  & -12.15 & \ms{0.13}{1.22}  & \ms{-0.18}{1.56} & 0.10 & -0.12 \\
        Bollinger Bands & 2.59  & -0.41  & \ms{0.04}{0.40}  & \ms{0.00}{0.58} & 0.11 & -0.01 \\
        \midrule
        GPT-3.5-turbo & 12.35 & -17.28 & \ms{0.25}{2.31} & \ms{-0.24}{2.55} & 0.09 & -0.12 \\ 
        GPT-4 & 22.68 & -15.61 & \ms{0.37}{2.11} & \ms{-0.23}{2.47} & 0.14 & -0.11 \\ 
        GPT-4o & 21.90 & -16.70 & \ms{0.36}{2.29} & \ms{-0.24}{2.81} & 0.15 & -0.12 \\ 
        o1-mini & 16.59 & -18.50 & \ms{0.30}{2.45} & \ms{-0.26}{2.41} & 0.12 & -0.13 \\ 
        \bottomrule
    \end{tabular}%
    }
    \label{tab:eth-selected}
    \vspace{-0.2cm}
\end{table}
\vspace{-0.2cm}
\end{minipage}
\end{wrapfigure}
\vspace{-0.2cm}
However, we observe that stronger LLMs sometimes underperform in trading scenarios, as noted in several studies \citep{li2024reflective, yu2024finmem}. Their LLM multi-agent frameworks based on stronger models (e.g., GPT-4-turbo) fail to align with the performance of weaker models (e.g., GPT-4, GPT-3.5-turbo). A similar phenomenon has been observed in studies on scientific discovery and medical domains \citep{chen2023huatuogpt, weng2024cycleresearcher}. Despite being a relatively common occurrence, no related research has explored this counterintuitive phenomenon in depth so far. 

To validate our observation, we conduct experiments using a single LLM instead of a multi-agent LLM system to eliminate potential biases from framework design. We evaluate LLMs including GPT-3.5-turbo, GPT-4, GPT-4o \citep{achiam2023gpt}, and o1-mini \citep{o1mini2024} on Bitcoin (BTC), Ethereum (ETH), and Solana (SOL) due to their popularity and significant market influence. The results for ETH, presented in \autoref{tab:eth-selected}, indicate that stronger LLMs (o1-mini, GPT-4o) do not always outperform weaker LLMs (GPT-4, GPT-3.5-turbo). Similar trends are observed in BTC and SOL results, detailed in \autoref{reimplement}. This unexpected finding motivates the following research questions:

\begin{quote}
\emph{Why stronger LLMs with advanced reasoning ability fail to outperform weaker ones in trading? How to better exploit their advanced reasoning ability?}
\end{quote}
To address these questions, we conduct an in-depth investigation into the reasoning processes of various LLMs, focusing on how they interpret news and make trading decisions. While previous approaches directly use news for analysis, we adopt a more fine-grained method by categorizing news into two distinct types: (1) factual, representing objective information such as events and data, and (2) subjective, reflecting personal opinions and judgments. This distinction is motivated by the significant influence of subjective judgments on cryptocurrency prices \citep{aggarwal2019understanding, anamika2022news, lee2023too}. To leverage this distinction, we introduce two specialized LLM agents: one to extract factual information and the other to extract subjective information. These agents independently analyze asset prices based on their respective components, and their insights are then integrated by another LLM agent, which considers the reasoning to provide a final trading decision.

\begin{figure}[ht]
\centering 
\includegraphics[width=0.9\linewidth]{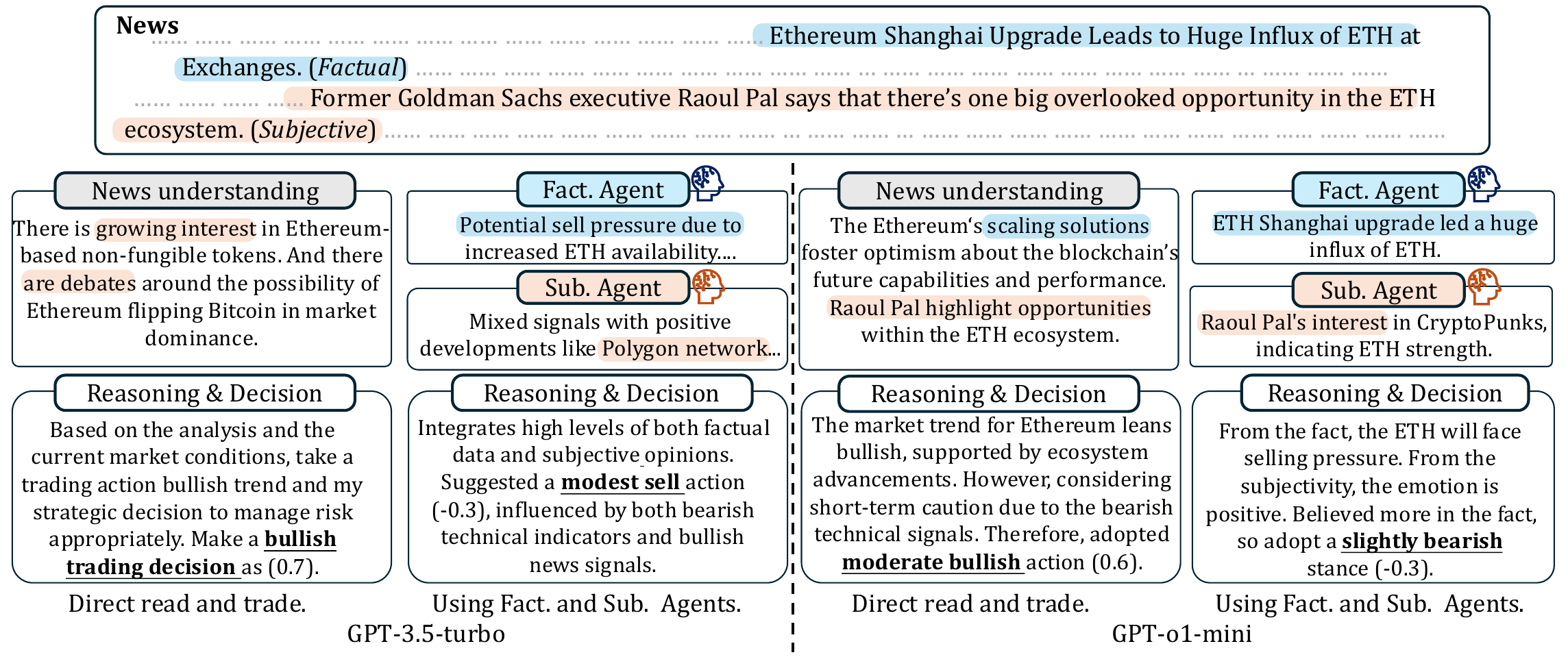} 
\caption{Comparison of Reasoning Processes - Trading Decisions Using News Data Alone; With/Without Fact and Subjectivity Agents on April 18, 2023 in the ETH Market, comparing GPT-3.5-turbo and o1-mini. The floating-point numbers represent buy/sell actions, where 0.7 indicates using 70\% of available cash to buy ETH, and -0.3 indicates selling 30\% of held ETH.}
\label{fig:reasoning} 
\end{figure}
We compare the traditional direct reasoning approach with our separate factual and subjective reasoning framework, as illustrated in \autoref{fig:reasoning}. In this case, the most profitable action is: selling all ETH it holds as as ETH's price on that date was \textit{the highest in the subsequent three months.} From the reasoning process comparison above, we draw two key insights:

\begin{itemize}[leftmargin=*]
\item \textbf{Stronger LLMs prioritize factual information.} In both using factual and subjective agents cases, GPT-o1-mini shows more belief in fact "Believed more in the fact". However, this focus on facts by stronger LLMs does not always lead to higher returns in cryptocurrency trading as economic theories suggesting that market participants are often influenced by emotional and psychological factors, driving asset prices beyond intrinsic values \citep{rubinstein2001rational, meltzer2002rational}.
\item \textbf{Splitting factual and subjective reasoning improves LLMs' profitability.}: The separated reasoning framework enables LLMs to make more profitable trading decisions. In this case of ETH's price on April 18, 2023, both GPT-3.5-turbo and o1-mini recommended more profitable actions under the split framework. This outcome reflects the splitting factual and subjective reasoning in news enhances LLM performance in trading scenarios. 
\end{itemize}

\begin{figure*}[ht]
\centering 
\includegraphics[width=0.75\linewidth]{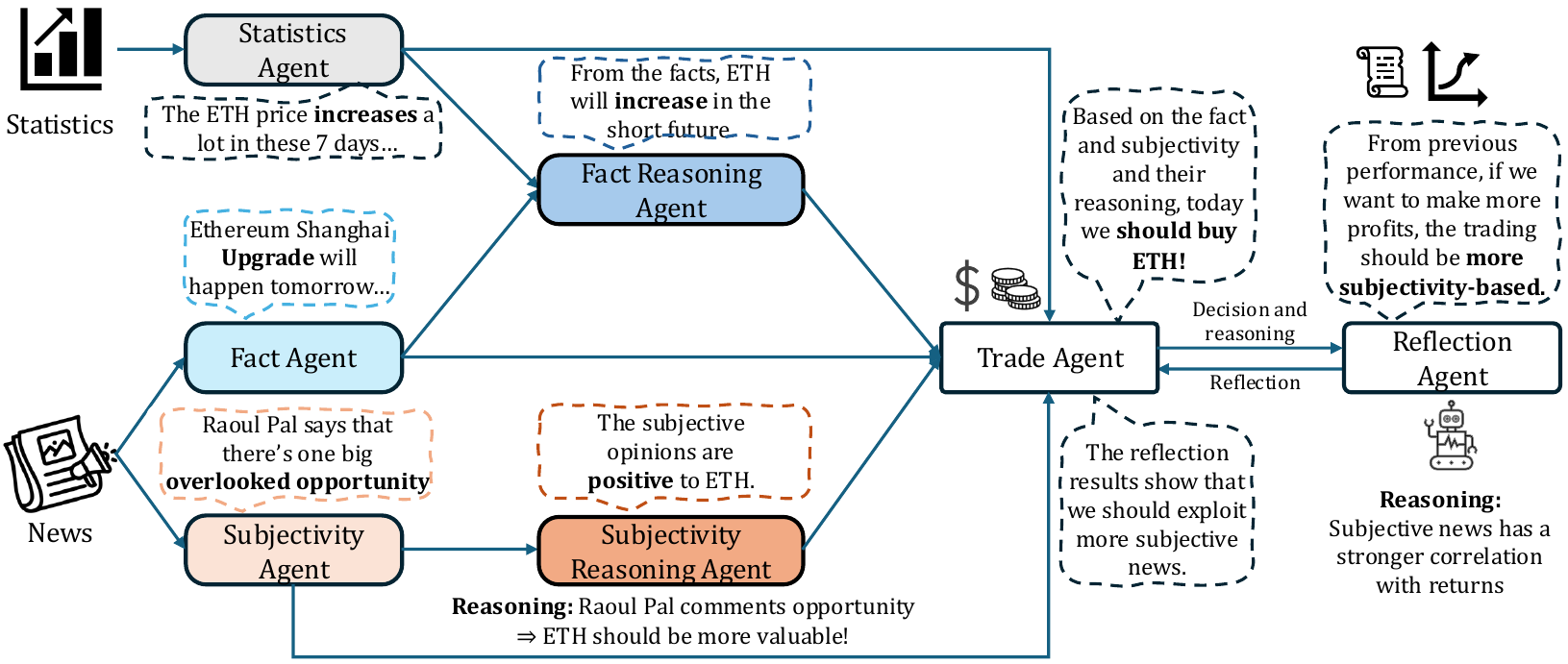} 
\caption{Fact-Subjectivity Reasoning Agent Framework. This framework contains the following agents: Statistics Agent, Fact Agent, Subjectivity Agent, Fact Reasoning Agent, Subjectivity Agent, Trade Agent, and Reflection Agent. We provide an example of each agent's analysis displayed besides the corresponding agent.} 
\label{fig:framework} 
\end{figure*}

Motivated by the above insights, we propose a novel multi-agent framework, Fact-Subjectivity-ReasoningAgent (FS-ReasoningAgent), which makes trading decisions by reasoning on both factual data and subjectivity. The framework is illustrated in \autoref{fig:framework}. FS-ReasoningAgent splits the reasoning process into a hierarchical structure through multiple agents: (1) dividing raw input data as statistics, factual and subjective news; (2) summarizing and reasoning according to factual or subjective information; (3) trading based on the processed information and reflection; (4) reflecting based on market returns, trading decisions and reasoning processes. \textit{FS-ReasoningAgent sets itself apart from previous LLM-based trading agents through its fine-grained reasoning, effectively balancing factual analysis with subjective interpretation for making more profitable decisions.}

To evaluate the performance of FS-ReasoningAgent in cryptocurrency trading, we conduct experiments on BTC, ETH, and SOL under both bull and bear market conditions between November 2023 and July 2024. The results show that our approach significantly outperforms CryptoTrade across all three cryptocurrencies in both bull and bear markets, achieving substantial increases in both returns and sharpe ratios. Moreover, FS-ReasoningAgent achieves results comparable to the traditional trading strategy - Buy and Hold. Furthermore, our ablation study of the FS-ReasoningAgent provides interesting insights: \textit{relying on subjective information leads to higher returns in bull markets, while focusing on factual data can result in better performance in bear markets.}

Our findings and contributions are as follows:
\begin{itemize}[leftmargin=*]
    \item  \textbf{Stronger LLMs Do Not Necessarily Outperform Weaker LLMs in Trading.} Our experiments reveal a counterintuitive phenomenon: stronger LLMs do not always outperform weaker LLMs in trading. This occurs because stronger LLMs show a preference for factual information over subjectivity. While this bias is beneficial in tasks like mathematics or coding, it can be less effective in emotion-driven trading markets.

    \item \textbf{Fact-Subjectivity-Aware Reasoning Multi-Agent Framework for Cryptocurrency Trading.}  
    FS-ReasoningAgent is a novel framework that separates factual and subjective information along with their corresponding reasoning processes. This design enables stronger LLMs to achieve higher trading profits than weaker LLMs by fully utilizing their advanced reasoning capabilities.

    \item \textbf{Empirical Validation and Insights.}  
    Experiments conducted across various cryptocurrencies and market conditions demonstrate that FS-ReasoningAgent is:  
    \textbf{(1) High-performing}: Achieving comparable results with traditional trading strategies and delivering over a 10\% performance increase compared to CryptoTrade in SOL trading.  
    \textbf{(2) Unlocking Advanced LLM Potential}: Models such as o1-mini and GPT-4o exhibit superior reasoning abilities in trading compared to GPT-3.5-turbo and GPT-4.  
    \textbf{(3) Providing Market Insights}: Subjective reasoning proves more critical in bull markets, while factual reasoning becomes essential in bear markets, offering valuable guidance for traders and researchers.
\end{itemize}

\section{FS-ReasoningAgent Framework}
In this section, we first provide the data collection process in our experiments and the FS-ReasoningAgent framework. Then, based on the experiment results, we analyze why stronger LLMs with advanced reasoning ability fail to outperform weaker ones. Then, built upon our analysis, we design the FS-ReasoningAgent framework as shown in \autoref{fig:framework}. 


\subsection{Data Collection}
We collect data from various open-source websites. The ethical requirements are explained in \autoref{collection.ethics}. Specifically, we obtain historical market statistics from CoinMarketCap—gathering daily data on prices, trading volumes, market capitalization, and other key metrics for BTC, ETH, and SOL—and we retrieve cryptocurrency-related news articles using the Gnews API, focusing on reputable sources such as Bloomberg, Yahoo Finance, and crypto.news to ensure comprehensive and diverse coverage. More details are in \autoref{collection.details}.

\subsection{Anaysis of LLM Reasoning in Trading}

In financial markets, news plays a critical role in shaping asset prices \citep{goldstein2023information, dhingra2024stock}. While stronger LLMs possess advanced reasoning abilities that lead them to prioritize factual information over subjective opinions in the news, this fact-driven approach may result in suboptimal trading decisions as economic theories suggest that market participants are often influenced by emotional and psychological factors, causing asset prices to deviate from their intrinsic values \citep{rubinstein2001rational, meltzer2002rational}. \textbf{Since trading markets are not entirely rational, LLM-based trading frameworks must adapt to this characteristic.}

Building on this insight, we introduce two specialized agents responsible for extracting factual and subjective components from the news. By delegating these tasks to separate agents, each agent can better focus on its specific extraction process. The trading agent then leverages the processed information from both agents, enabling more comprehensive and balanced trading decisions. We present the detailed agent design in the following sections.



\subsection{Component Design of FS-ReasoningAgent}
After data collection and analyzing LLM reasoning in trading, we introduce each component of the FS-ReasoningAgent, demonstrating how the framework makes its trading decisions, as illustrated in \autoref{fig:framework}.
\textbf{Statistics Agent.} Statistics Agent is responsible for extracting, analyzing, and summarizing key market data related to cryptocurrencies. It reads various quantitative metrics such as the opening price, total transaction volume, average gas fees, unique addresses, and total value transferred on the cryptocurrency. Based on this data, the Statistics Agent identifies short-term market trends and provides an essential foundation for the overall trading strategy. This agent plays a vital role in ensuring that trading decisions are grounded in up-to-date, quantifiable market conditions. An example of Statistics Agent is shown in \autoref{table:statAgent}.

\begin{table*}[ht]
\centering
\fontsize{9}{9}\selectfont 
\caption{An example of Statistics Agent.}
\vspace{-0.3cm}
\begin{tabular}{p{\textwidth}} 
\toprule[1.5pt] 
\midrule 
\textbf{Prompts:} \\
You are an eth statistics agent. The recent price and auxiliary information is given in chronological order below: \\
\{Open price: 2241.75, unique addresses: 577757, ...; Open price: 2317.97, unique addresses: 576510, ...\} \\
Write one concise paragraph to analyze the recent information and estimate the statistical trend accordingly. \\
\midrule 
\textbf{Responses:} \\
Over the recent period, Ethereum's open price has demonstrated a notable upward trend...towards continued positive momentum in Ethereum's market performance. \\
\midrule 
\bottomrule[1.5pt]
\label{table:statAgent}
\end{tabular}
\end{table*}

\textbf{Fact Agent.} Fact Agent focuses on gathering and analyzing factual news related to the cryptocurrency market. It filters out subjective commentary, relying instead on concrete events such as regulatory updates, technological advancements, and major market shifts. The Fact Agent improves the trading decision process by identifying impactful facts, such as Ethereum’s technological progress and regulatory updates related to Ethereum ETFs shown in the news. This information is crucial for generating rational trading strategies, as it provides context on real-world factors that can influence the market. An example of Fact Agent is shown in \autoref{table:factAgent}.

\begin{table*}[ht]
\centering
\fontsize{9}{9}\selectfont 
\caption{An example of Fact Agent.}
\vspace{-0.3cm}
\begin{tabular}{p{\textwidth}} 
\toprule[1.5pt] 
\midrule 
\textbf{Prompts:}
You are an eth fact agent. You are required to analyze only the factual news, not the subjective news such as someone's comments from following news articles:\\
\{'title': 'XRP, Bitcoin and Ethereum Eye Unusual Transfer Activity Worth Millions...\} \\
Write one concise paragraph to analyze and summarize the factual news and estimate the market trend accordingly.\\
\midrule 
\textbf{Responses:} Recent significant movements in the Ethereum ecosystem highlight sustained investor activity...positive trend of ETH.\\
\midrule 
\bottomrule[1.5pt]
\label{table:factAgent}
\end{tabular}
\end{table*}


\textbf{Subjectivity Agent.} Subjectivity Agent is tasked with extracting and analyzing subjective opinions, sentiments, and expert commentary from news sources. This agent gathers market rumors, social media trends, and expert predictions that reflect emotions and biases in the market. By capturing market sentiment, this agent helps identify how emotional responses may drive short-term price volatility and offers insights into potential market shifts driven by psychological factors. An example of a subjectivity agent is shown in \autoref{table:subjectAgent}.

\begin{table*}[ht]
\centering
\fontsize{9}{9}\selectfont 
\vspace{-0.3cm}
\caption{An example of Subjectivity Agent.}
\vspace{-0.3cm}
\begin{tabular}{p{\textwidth}} 
\toprule[1.5pt] 
\midrule 
\textbf{Prompts:}
You are an eth trading analyst. You are required to analyze only the subjective news such as someone's comments from following news articles:\\
\{'title': 'Analyst projects Ethereum potentially to 2k...\} \\
Write one concise paragraph to analyze the subjective news and estimate the market trend accordingly.\\
\midrule 
\textbf{Responses:} FxPro senior analyst Alex Kuptsikevich forecasts a potential decline of ETH to 2k...ETH will appear downward.\\
\midrule 
\bottomrule[1.5pt]
\label{table:subjectAgent}
\end{tabular}
\end{table*}


\textbf{Fact Reasoning Agent.} Fact Reasoning Agent synthesizes information from both the Statistics Agent and the Fact Agent to provide logical, data-driven reasoning about current market conditions. This agent ensures that the trading strategy is grounded in objective data. This agent's role is to ensure that the factual information is processed to generate actionable insights for trading decisions. An example of Fact Reasoning Agent is shown in \autoref{table:factReasonAgent}.

\begin{table*}[ht]
\centering
\fontsize{9}{9}\selectfont 
\caption{An example of Fact Reasoning Agent.}
\vspace{-0.3cm}
\begin{tabular}{p{\textwidth}} 
\toprule[1.5pt] 
\midrule 
\textbf{Prompts:}
You are an eth fact reasoning agent. Based on the following factual news analysis and statistics analysis, provide your reasoning for the trading suggestions. This reasoning will be used for the final trading action.\\
Factual News Analysis: \{Fact Agent Responses\} \\
Statistics Analysis: \{Statistics Agent Responses\}\\
\midrule 
\textbf{Responses:} The following factors: Liquidity Influx, Technological Advancements, ...ETH exhibits positive growth trajectory.\\
\midrule 
\bottomrule[1.5pt]
\label{table:factReasonAgent}
\end{tabular}
\end{table*}


\textbf{Subjectivity Reasoning Agent.} Subjectivity Reasoning Agent interprets the subjective insights gathered by the Subjectivity Agent, offering a more fine-grained analysis on market trends. This agent considers how emotions, biases, and opinions may influence market movements and price volatility. By reasoning in these subjective elements, this agent provides a complementary layer of reasoning to fact-based analysis, enriching the overall decision-making process. An example of Subjectivity Reasoning Agent is shown in \autoref{table:subjectReasonAgent}.

\begin{table*}[ht]
\centering
\fontsize{9}{9}\selectfont 
\caption{An example of Subjectivity Reasoning Agent.}
\vspace{-0.3cm}
\begin{tabular}{p{\textwidth}} 
\toprule[1.5pt] 
\midrule 
\textbf{Prompts:}
You are an eth subjectivity reasoning agent. Based on the following subjective news summary and analysis, provide your reasoning for the trading suggestions. This reasoning will be used for the final trading action.\\
Subjective News Analysis: \{Subjectivity Agent Responses.\}\\
\midrule 
\textbf{Responses:} Given influencers highlighting ETH vulnerability...immediate market conditions warrant a risk-managed approach. \\
\midrule 
\bottomrule[1.5pt]
\label{table:subjectReasonAgent}
\end{tabular}
\end{table*}


\textbf{Trade Agent.} Trade Agent serves as the decision-making core of the FS-ReasoningAgent framework, synthesizing inputs from the Statistics Agent, Fact Agent, Fact Reasoning Agent, and Subjectivity Reasoning Agent to make final trading decisions. Converts the collective analysis into an actionable decision, represented on a continuous scale from $\left[-1, 1\right]$, where \(-1\) means a full sell action, \(0\) represents a hold, and \(1\) indicates a full buy action. The design of assigning buy/sell decisions and their corresponding percentages to the LLM is inspired by common practices in human trading as it is standard for traders to determine not only whether to buy or sell but also how much of their portfolio to allocate to a particular action \citep{jang2023deep, cui2024multi}. Trade Agent carefully balances factual data and subjective sentiment to optimize trades for profit while managing risk. Upon executing a trade, a proportional transaction fee is applied based on the value traded. An example of Trade Agent is in \autoref{table:tradeAgent}.

\begin{table*}[ht]
\centering
\fontsize{9}{9}\selectfont 
\caption{An example of Trade Agent.}
\vspace{-0.3cm}
\begin{tabular}{p{\textwidth}} 
\toprule[1.5pt] 
\midrule 
\textbf{Prompts:}
You are an experienced eth trader and you are trying to maximize your overall profit by trading eth. In each day, you must make an action to buy or sell eth. You are assisted by a few agents below and need to decide the final action.\\
STATISTICS AGENT REPORT: "\{REPORT.\}" \\
... \\
REFLECTION AGENT REPORT: "\{REPORT.\}" \\
Now, provide your response in the following format:\\
1. Reasoning: Briefly analyze the given reports...factual and subjective elements.\\        
2. Factual vs Subjective Weighting: If there's a conflict between factual and subjective information, explain which you favor and why. \\  
3. Risk Management: Describe how you're managing risk.\\
4. Action: Indicate your trading action as a 1-decimal float in the range of [-1,1]. \\        
\midrule 
\textbf{Responses:} Action: -0.4...Slight sell to reduce exposure while acknowledging underlying network strength and current bearish sentiment.\\
\midrule 
\bottomrule[1.5pt]
\label{table:tradeAgent}
\end{tabular}
\end{table*}



\textbf{Reflection Agent.} Reflection Agent plays a critical role in learning and adapting the FS-ReasoningAgent’s trading strategy over time. It reviews past trading actions and outcomes, analyzing the effectiveness of the reasoning process and the information used in decision-making. By examining recent prompts, decisions, and market returns, the Reflection Agent identifies which types of information—factual data or subjective opinions—had the most significant impact on trading success. This feedback loop allows the system to adjust future strategies, improving performance by focusing on the most influential factors. An example of this reflective process is illustrated in \autoref{table:reflectionAgent}.

\begin{table*}[ht]
\centering
\fontsize{9}{9}\selectfont 
\caption{An example of Reflection Agent.}
\vspace{-0.3cm}
\begin{tabular}{p{\textwidth}} 
\toprule[1.5pt] 
\midrule 
\textbf{Prompts:} You are an eth reflection agent. Reflect on your recent trading performance and provide guidance for future trades: ... \\
\midrule 
\textbf{Responses:} To maximize trading performance in the current Ethereum market conditions, maintain a balanced approach with approximately 60\% weighting on factual information and 40\% on subjectivity... \\
\midrule 
\bottomrule[1.5pt]
\label{table:reflectionAgent}
\end{tabular}
\end{table*}
\section{Experiments}

In this section, we detail the experiments designed to evaluate the performance of FS-ReasoningAgent in comparison to established baseline strategies in the cryptocurrency trading domain.

\subsection{Experimental Setup}


\begin{table*}[h!]
    \centering
\caption{Dataset splits with prices in US dollars. Each split includes the start date but excludes the end date for transaction days. The total profit is evaluated on the end date.} \label{tab:period-stats}
\resizebox{0.7\linewidth}{!}{
\begin{tabular}{ccccccc}
\toprule
Type                 & Split           & Start      & End        & Open     & Close    & Trend    \\ 
\midrule
\multirow{3}{*}{BTC} & Validation      & 2023-11-16 & 2024-01-15 & 37879.97 & 42511.96 & \textcolor{mhold}{12.23\%}  \\
                     & Test   Bullish  & 2024-01-24 & 2024-03-13 & 39877.59 & 71631.35 & \textcolor{minc}{79.63\%}  \\ 
                     & Test   Bearish  & 2024-05-21 & 2024-07-13 & 71443.06 & 59231.95 & \textcolor{mdec}{-17.09\%} \\ \hline
\multirow{3}{*}{ETH} & Validation      & 2023-11-10 & 2024-01-08 & 2121.06  & 2333.03  & \textcolor{mhold}{9.99\%}   \\
                     & Test   Bullish  & 2024-01-24 & 2024-03-13 & 2241.74  & 4006.45  & \textcolor{minc}{78.72\%}  \\ 
                     & Test   Bearish  & 2024-05-27 & 2024-07-08 & 3826.13  & 2929.86 & \textcolor{mdec}{-23.42\%} \\ \hline
\multirow{3}{*}{SOL} & Validation      & 2023-11-16 & 2024-01-08 & 65.53   & 97.79    & \textcolor{mhold}{49.18\%}  \\
                     & Test   Bullish  & 2024-01-24 & 2024-03-13 & 84.28    & 151.02    & \textcolor{minc}{77.35\%} \\ 
                     & Test   Bearish  & 2024-05-21 & 2024-07-11 & 186.51   & 127.61    & \textcolor{mdec}{-15.53\%} \\
\bottomrule
\end{tabular}
}
\vspace{-0.1cm}
\end{table*}
\vspace{-0.1cm}


\stitle{Datasets.} To ensure our experiments are robust across different cryptocurrencies and market conditions, we base our study on a dataset covering several months, detailed in \autoref{tab:period-stats}. This dataset captures the recent market performance of BTC, ETH, and SOL, highlighting challenges in identifying market trends and volatility. We divide the dataset into validation and test sets, using the validation set to fine-tune model hyperparameters and prompts, and the test set to evaluate model performance. The data period, spanning from November 2023 to July 2024, is carefully chosen to prevent data leakage, as all GPT models have a knowledge cutoff prior to November 2023\footnote{\url{https://openai.com/api/pricing/}}. The dataset covers both bull and bear markets, allowing us to assess the effectiveness of both the baseline models and our proposed model \citep{baroiu2023bitcoin, cagan2024stock, li2024reflective}.

\begin{wrapfigure}{R}{0.5\textwidth}
\begin{minipage}{0.5\textwidth}
\begin{table}[H]
\centering
\caption{\small Performance of each strategy on BTC under both bull and bear market conditions. For each market condition and metric, the best result is highlighted in bold, the runner-up is indicated with an underline, and the best result among each families of LLM-based strategies is highlighted in green.}
\small 
    \resizebox{1.0\textwidth}{!}{%
    \begin{tabular}{l|cc|cc|cc}
        \toprule
        \textbf{Strategy} & \multicolumn{2}{c|}{\textbf{Total Return}} & \multicolumn{2}{c|}{\textbf{Daily Return}} & \multicolumn{2}{c}{\textbf{Sharpe Ratio}} \\
        \cmidrule(lr){2-3} \cmidrule(lr){4-5} \cmidrule(lr){6-7}
        & \textbf{Bull} & \textbf{Bear} & \textbf{Bull} & \textbf{Bear} & \textbf{Bull} & \textbf{Bear} \\
        \midrule
        Buy and Hold & \textbf{79.63} & -19.15 & \ms{1.18}{2.21} & \ms{-0.38}{1.79} & \textbf{0.53} & -0.21 \\
        SMA & 69.51 & -9.80 & \ms{1.09}{2.57} & \ms{-0.19}{0.76} & 0.43 & -0.25 \\
        SLMA & 53.09 & \underline{-8.30} & \ms{0.89}{2.49} & \ms{\underline{-0.16}}{0.97} & 0.36 & \underline{-0.16} \\
        MACD & 22.01 & -15.26 & \ms{0.41}{1.28} & \ms{-0.29}{1.66} & 0.32 & -0.18 \\
        Bolling Bands & 8.28 & \textbf{-6.10} & \ms{0.16}{0.51} & \ms{\textbf{-0.11}}{1.01} & 0.32 & \textbf{-0.11} \\
        \midrule
        CryptoTrade(GPT-3.5-turbo) & {70.25} & \textcolor{minc}{-18.08} & \ms{1.12}{2.53} & \ms{\textcolor{minc}{-0.36}}{1.75} & \textcolor{minc}{0.44} & \textcolor{minc}{-0.21} \\
        CryptoTrade(GPT-4) & 66.83 & -21.11 & \ms{1.08}{2.21} & \ms{-0.43}{1.66} & 0.39 & -0.26 \\
        CryptoTrade(GPT-4o) & 68.35 & -20.21 & \ms{1.10}{2.57} & \ms{-0.41}{1.68} & 0.43 & -0.24 \\
        CryptoTrade(o1-mini) & \textcolor{minc}{70.83} & -19.89 & \ms{\textcolor{minc}{1.13}}{2.58} & \ms{-0.40}{1.61} & \textcolor{minc}{0.44} & -0.25 \\
        \midrule
        Ours(GPT-3.5-turbo) & 73.55 & -19.15 & \ms{1.16}{2.61} & \ms{-0.39}{1.71} & 0.23 & \textcolor{minc}{-0.23} \\
        Ours(GPT-4) & \underline{\textcolor{minc}{77.47}} & -15.23 & \ms{\textbf{\textcolor{minc}{1.21}}}{2.63} & \ms{-0.30}{1.20} & \underline{\textcolor{minc}{0.46}}& -0.25 \\
        Ours(GPT-4o) & 74.27 & \textcolor{minc}{-13.94} & \ms{1.17}{2.60} & \ms{\textcolor{minc}{-0.28}}{0.85} & 0.45 & -0.33 \\
        Ours(o1-mini) & 76.19 & -15.91 & \ms{\underline{1.20}}{2.62} & \ms{-0.32}{0.93} & \underline{\textcolor{minc}{0.46}} & -0.35 \\
        \bottomrule
    \end{tabular}
    }
\label{tab:btc}
\end{table}
\end{minipage}
\end{wrapfigure}

\stitle{Evaluation Metrics.} We initialize the FS-ReasoningAgent with a starting capital of one million US dollars, evenly split between cash and BTC/ETH/SOL, allowing it to capitalize on both buying and selling opportunities in the cryptocurrency market. At the end of the trading session, we assess performance using the following commonly accepted metrics: \textbf{Return}, \textbf{Sharpe Ratio}, \textbf{Daily Return Mean}, and \textbf{Daily Return Std}. This evaluation approach ensures a thorough and unbiased comparison between FS-ReasoningAgent and baseline strategies. Details are in \autoref{metrics}.





\stitle{Baseline Strategies.} To benchmark FS-ReasoningAgent's performance, we compare it against widely recognized baseline trading strategies. The baselines are detailed in \autoref{baselines}.

\subsection{Experimental Results}
The performance comparison between various trading strategies and FS-ReasoningAgent is presented in \autoref{tab:btc}, \autoref{tab:eth}, and \autoref{tab:sol}. Key findings are as follows:

\begin{wrapfigure}{R}{0.5\textwidth}
\begin{minipage}{0.5\textwidth}
\begin{table}[H]
    \centering
        \caption{\small Performance of each strategy on ETH under bull and bear market conditions.}
    \small 
    \resizebox{1.0\textwidth}{!}{%
    \begin{tabular}{l|cc|cc|cc}
        \toprule
        \textbf{Strategy} & \multicolumn{2}{c|}{\textbf{Total Return (\%)}} & \multicolumn{2}{c|}{\textbf{Daily Return (\%)}} & \multicolumn{2}{c}{\textbf{Sharpe Ratio}} \\
        \cmidrule(lr){2-3} \cmidrule(lr){4-5} \cmidrule(lr){6-7}
        & \textbf{Bull} & \textbf{Bear} & \textbf{Bull} & \textbf{Bear} & \textbf{Bull} & \textbf{Bear} \\
        \midrule
        Buy and Hold & \textbf{78.72} & -23.63 & \ms{1.18}{2.21} & \ms{-0.60}{2.13} & \underline{0.53} & -0.28 \\
        SMA & 59.60 & \underline{-19.13} & \ms{0.96}{2.11} & \ms{\underline{-0.49}}{1.00} & 0.45 & -0.49 \\
        SLMA & 60.31 & \textbf{-9.01} & \ms{0.97}{2.07} & \ms{\textbf{-0.21}}{1.34} & 0.47& \textbf{-0.16} \\
        MACD & 12.93 & -20.10 & \ms{0.25}{0.78} & \ms{-0.50}{2.00} & 0.32 & \underline{-0.25} \\
        Bollinger Bands & 77.24 & -23.68 & \ms{1.17}{2.20} & \ms{-0.60}{2.12} & \underline{0.53} & {-0.29} \\
        \midrule
        CryptoTrade(GPT-3.5-turbo) & {74.83} & \textcolor{minc}{-22.35} & \ms{{1.17}}{2.20} & \ms{\textcolor{minc}{-0.58}}{1.94} & \textcolor{minc}{0.53} & -0.30 \\
        CryptoTrade(GPT-4) & 74.41 & -23.06 & \ms{1.17}{2.20} & \ms{-0.60}{2.08} & \textcolor{minc}{0.53} & \textcolor{minc}{-0.29} \\
        CryptoTrade(GPT-4o) & 74.23 & -24.13 & \ms{1.16}{2.18} & \ms{-0.63}{2.11} & \textcolor{minc}{0.53} & -0.30 \\
        CryptoTrade(o1-mini) & \textcolor{minc}{75.01} & -23.68 & \ms{\textcolor{minc}{1.17}}{2.19} & \ms{-0.62}{2.15} & \textcolor{minc}{0.53} & \textcolor{minc}{-0.29} \\
        \midrule
        Ours(GPT-3.5-turbo) & 71.09 & -22.33 & \ms{1.12}{2.15} & \ms{-0.58}{1.92} & 0.52 & \textcolor{minc}{-0.30} \\
        Ours(GPT-4) & 76.67 & -23.41 & \ms{\underline{1.19}}{2.21} & \ms{-0.61}{1.97} & \textbf{\textcolor{minc}{0.54}} & -0.31 \\
        Ours(GPT-4o) & 76.74 & -21.64 & \ms{\underline{1.19}}{2.21} & \ms{\textcolor{minc}{-0.56}}{1.82} & \textbf{\textcolor{minc}{0.54}} & -0.31 \\
        Ours(o1-mini) & \underline{\textcolor{minc}{77.28}} & \textcolor{minc}{-21.88} & \ms{\textcolor{minc}{\textbf{1.20}}}{2.22} & \ms{-0.57}{1.79} & \textbf{\textcolor{minc}{0.54}} & -0.32 \\
        \bottomrule
    \end{tabular}%
    }
        \label{tab:eth}      
\end{table}
\end{minipage}
\end{wrapfigure}

\stitle{Finding 1: FS-ReasoningAgent's Comparable Performance with Traditional Trading Strategies.} FS-ReasoningAgent performs competitively against traditional trading strategies across diverse market conditions. In bullish markets, it consistently ranks among the top performers, achieving a 77.28\% total return with a Sharpe ratio of 0.54 in the ETH market, surpassing most traditional strategies. In bearish markets, FS-ReasoningAgent effectively reduces losses, such as limiting losses to -14.52\% in the SOL market, outperforming methods like SMA (-27.17\%) and MACD (-15.44\%). Its fact-subjective reasoning mechanism enables adaptive trading behavior, making it a strong alternative to established trading approaches.

\stitle{Finding 2: FS-ReasoningAgent Improves LLMs' Trading Capabilities. }FS-ReasoningAgent consistently outperforms CryptoTrade across BTC, ETH, and SOL in both bull and bear markets. For BTC, FS-ReasoningAgent (GPT-4) achieves a 77.47\% return in a bull market, surpassing CryptoTrade’s best-performing model (GPT-3.5-turbo) by 7\%, while limiting bear market losses to -13.94\%, compared to CryptoTrade’s -20.21\%. Similarly, in SOL’s bull market, FS-ReasoningAgent (o1-mini) delivers a 76.71\% return, over 10\% higher than CryptoTrade’s GPT-3.5-turbo at 66.64\%. FS-ReasoningAgent also achieves better Sharpe ratios, indicating improved risk-adjusted returns. These results highlight that splitting factual and subjective reasoning is an effective approach, enabling FS-ReasoningAgent to serve as a robust trading solution in volatile market conditions.

\begin{wrapfigure}{R}{0.5\textwidth}
\begin{minipage}{0.5\textwidth}
\begin{table}[H]
    \centering
    \caption{\small Performance of each strategy on SOL under bull and bear market conditions.}
    \small 
    \resizebox{1.0\textwidth}{!}{%
    \begin{tabular}{l|cc|cc|cc}
        \toprule
        \textbf{Strategy} & \multicolumn{2}{c|}{\textbf{Total Return (\%)}} & \multicolumn{2}{c|}{\textbf{Daily Return (\%)}} & \multicolumn{2}{c}{\textbf{Sharpe Ratio}} \\
        \cmidrule(lr){2-3} \cmidrule(lr){4-5} \cmidrule(lr){6-7}
        & \textbf{Bull} & \textbf{Bear} & \textbf{Bull} & \textbf{Bear} & \textbf{Bull} & \textbf{Bear} \\
        \midrule
        Buy and Hold & \textbf{77.35} & -24.08 & \ms{\textbf{1.23}}{3.39} & \ms{-0.45}{3.97} & \textbf{0.36} & -0.11 \\
        SMA & 42.09 & -27.17 & \ms{0.74}{2.65} & \ms{-0.58}{2.37} & 0.28 & -0.24 \\
        SLMA &47.84  & -18.92 & \ms{0.83}{2.93} & \ms{-0.39}{1.74} & 0.28  & -0.22 \\
        MACD & 34.63 & -15.44 & \ms{0.62}{2.17} & \ms{-0.29}{2.58} &  0.29& -0.11 \\
        Bollinger Bands &22.97  & \textbf{-8.94} & \ms{0.42}{1.23} & \ms{\textbf{-0.13}}{3.15} & 0.34 & \textbf{-0.04} \\
        \midrule

        CryptoTrade(GPT-3.5-turbo)          & \textcolor{minc}{66.64} & -23.56 & \ms{\textcolor{minc}{1.10}}{3.25} & \ms{-0.45}{3.77} & \textcolor{minc}{0.34} & -0.12 \\
        CryptoTrade(GPT-4)       & 32.59 & \textcolor{minc}{-21.51} & \ms{0.61}{2.65} & \ms{\textcolor{minc}{-0.41}}{3.65} & 0.23 & \textcolor{minc}{-0.11} \\
        CryptoTrade(GPT-4o)     & 48.41  & -24.63 & \ms{0.84}{2.52} & \ms{-0.48}{3.83} & 0.33 & -0.13 \\
        CryptoTrade(o1-mini)       & 42.48  & -21.95 & \ms{0.76}{2.60} & \ms{-0.43}{3.40} & 0.29& -0.13 \\ \midrule
        Ours(GPT-3.5-turbo)   & 68.03 &-24.67 & \ms{1.12}{3.27} & \ms{-0.49}{3.55} & 0.34 & -0.14 \\ 
        Ours(GPT-4)   & 64.35 & -25.33 & \ms{1.07}{3.25} & \ms{-0.52}{3.07} & 0.33 & -0.16 \\ 
        Ours(GPT-4o)   & 69.67 & \underline{\textcolor{minc}{-14.52}} & \ms{1.14}{3.30} & \ms{\underline{\textcolor{minc}{-0.26}}}{3.05} & \underline{0.35} & \textcolor{minc}{\underline{-0.09}} \\ 
        Ours(o1-mini) & \underline{\textcolor{minc}{76.71}} & -19.40 & \ms{\underline{\textcolor{minc}{1.22}}}{3.38} & \ms{-0.36}{3.40} & \textbf{\textcolor{minc}{0.36}} & -0.11 \\ 
        \bottomrule
    \end{tabular}%
    }
\label{tab:sol}
\end{table}
\end{minipage}
\end{wrapfigure}
\stitle{Finding 3: FS-ReasoningAgent Makes Stronger LLMs Great Again.} The experimental results demonstrate that stronger LLMs, such as GPT-4 and o1-mini, achieve superior performance within the FS-ReasoningAgent framework due to its fact-subjectivity splitting mechanism. This structured reasoning process enables stronger LLMs to separate factual analysis from subjective interpretation, leading to more accurate and informed trading decisions. In contrast, without this separation, stronger LLMs often struggle, as evidenced by CryptoTrade's results, where GPT-4o underperforms GPT-3.5-turbo by 18\% in the SOL bull market. FS-ReasoningAgent highlights that unlocking the full potential of stronger LLMs requires an architectural design that harnesses their advanced reasoning capabilities through task-specific reasoning separation, resulting in better returns and reduced trading risks.

\subsection{Ablation Study}


\begin{wrapfigure}{R}{0.5\textwidth}
\begin{minipage}{0.5\textwidth}
\vspace{-0.2cm}
\begin{table}[H]
\centering
    \caption{\small FS-ReasoningAgent Ablation study of each agent's performance under BTC bull and bear market conditions.}
    \small
    \resizebox{\columnwidth}{!}{ 
    \begin{tabular}{l|cc|cc}
        \toprule
        \textbf{Components} & \multicolumn{2}{c|}{\textbf{Return (\%)}} & \multicolumn{2}{c}{\textbf{Sharpe Ratio}} \\
        \cmidrule(lr){2-3} \cmidrule(lr){4-5}
        & \textbf{Bull} & \textbf{Bear} & \textbf{Bull} & \textbf{Bear} \\
        \midrule
        Full                        & 76.19 & -15.91 & 0.46 & -0.35 \\
        w/o Reflection Agent        & 71.77 & -17.85 & 0.44 & -0.40 \\
        w/o Fact Reasoning Agent    & 72.23 & -19.21 & 0.43 & -0.39 \\
        w/o Sub. Reasoning Agent    & 66.04 & -16.83 & 0.42 & -0.36 \\
        w/o Statistics Agent        & 74.25 & -20.40 & 0.45 & -0.36 \\
        \bottomrule
    \end{tabular}
    }
    \label{tab:ablation_bull_bear}
\end{table}\end{minipage}
\end{wrapfigure}

To assess the contribution of each agent in the FS-ReasoningAgent framework, we conduct an ablation study using the o1-mini backbone on BTC under both bull and bear market conditions. In each iteration, we remove one component from the full framework. The results are in \autoref{tab:ablation_bull_bear}. Additionally, to compare the reasoning capabilities of FS-ReasoningAgent with CryptoTrade, we perform ablation studies on both frameworks to highlight the standalone impact of the reasoning mechanism by removing Reflection Agent component. The results are presented in \autoref{tab:comparison_with_arrows}. Based on these ablation studies, we have three insights:



\stitle{Insight 1: FS-ReasoningAgent significantly enhances LLM reasoning abilities for trading.}
When Reflection Agent is removed, CryptoTrade's performance declines significantly more than FS-ReasoningAgent's, as shown in \autoref{tab:comparison_with_arrows}. Since both frameworks utilize o1-mini as the backbone, this demonstrates that FS-ReasoningAgent enhances LLMs' standalone reasoning capabilities for trading, even without the reflection mechanism.

\stitle{Insight 2: Subjectivity is more important in the bull market.} The performance in the bull market, reflected by both returns and the sharpe ratio, suggests that subjective reasoning plays a crucial role in capturing the market's positive sentiment. Removing the Subjective Reasoning Agent results in a notable drop in returns from 76.19\% to 66.04\%, along with the largest decline in the sharpe ratio from 0.46 to 0.42. This indicates that in bullish markets, understanding and interpreting market sentiment—such as reactions to news, emotions-is essential for maximizing profits. 

The likely explanation is that during bull markets, price movements are often driven by investors' positive sentiment, which typically emerges earlier than factual changes.


\begin{wrapfigure}{R}{0.4\textwidth}
\begin{minipage}{0.4\textwidth}
\begin{table}[H]
\centering
\caption{Performance comparison of CryptoTrade and FS-ReasoningAgent, with decreases indicated by \textcolor{mdec}{$\downarrow$}.}
\resizebox{\columnwidth}{!}{%
\begin{tabular}{l|cc|cc}
\toprule
\textbf{Components} & \multicolumn{2}{c|}{\textbf{Return (\%)}} & \multicolumn{2}{c}{\textbf{Sharpe Ratio}} \\ 
\cmidrule(lr){2-3} \cmidrule(lr){4-5}
 & \textbf{Bull} & \textbf{Bear} & \textbf{Bull} & \textbf{Bear} \\ 
\midrule
Full (CryptoTrade) & 70.83 & -19.89 & 0.44 & -0.25 \\ 
w/o Reflection Agent & 59.87 & -24.33 & 0.35 & -0.32 \\ 
Decrease & \textcolor{mdec}{$\downarrow$10.96} & \textcolor{mdec}{$\downarrow$4.44} & \textcolor{mdec}{$\downarrow$0.09} & \textcolor{mdec}{$\downarrow$0.07} \\ 
\midrule
Full (FS-Reasoning) & 76.19 & -15.91 & 0.46 & -0.35 \\ 
w/o Reflection Agent & 71.77 & -17.85 & 0.44 & -0.40 \\ 
Decrease & \textcolor{mdec}{$\downarrow$4.42} & \textcolor{mdec}{$\downarrow$1.94} & \textcolor{mdec}{$\downarrow$0.02} & \textcolor{mdec}{$\downarrow$0.05} \\ 
\bottomrule
\end{tabular}%
}
\label{tab:comparison_with_arrows}
\end{table}\end{minipage}
\end{wrapfigure}
\stitle{Insight 3: Facts are more important in the bear market.} In bear markets, factual reasoning plays a critical role in minimizing losses. The study shows that removing the Fact Reasoning Agent leads to a deeper negative return of -19.21\%, compared to -15.91\% for the full framework. Similarly, the sharpe ratio drops from -0.35 to -0.39 without the factual component. A similar pattern is observed when the Statistics Agent is removed, causing the largest decrease in returns from -15.91\% to -20.40\%, as statistical data also represent factual insights. This highlights the importance of relying on clear data and objective analysis during bearish periods, when fear and pessimism dominate. 

The possible reason is that in bear markets, emotional reactions to market downturns can trigger irrational decisions, while fact-driven analysis helps maintain objectivity and reduce panic-driven trades. This aligns with the famous quote: \emph{"Be greedy when others are fearful."}

\section{Related Work}
\vspace{-0.1cm}
\noindent \textbf{LLMs for Trading Decisions.} 
Recent progress in LLMs has had a notable impact on economics and financial decision-making. Models specifically designed for finance, such as FinGPT, BloombergGPT, FinMA, FinAgent, FinMem \citep{liu2023fingpt, wu2023bloomberggpt, xie2023pixiu, zhang2024multimodal, yu2024finmem}, have been applied to tasks like sentiment analysis, entity recognition, and making trading decisions. LLM-driven agents for financial trading have also drawn considerable attention. The Sociodojo framework \citep{cheng2024sociodojo}, for instance, developed analytical agents for managing stock portfolios, demonstrating the potential for creating "hyper-portfolios." Although numerous studies focus on trading, few explore the performance differences between various LLM backbones in depth. For example, in the FinMem Backbone Algorithm Comparison \citep{yu2024finmem}, GPT-4-Turbo achieved a cumulative return that was less than 8\% of GPT-4's performance, a surprising result that warrants deeper analysis.

\noindent \textbf{Reasoning Process of LLM Agents.}
A common method for examining the reasoning process of LLMs involves generating intermediate reasoning steps using techniques such as chain-of-thought reasoning \citep{wei2022chain, kojima2022large} and question decomposition \citep{zhou2022least}. However, the reasoning process behind LLMs' trading decisions has been largely unexplored \citep{ding2024large, zhang2024llm}. To address this gap, we propose a FS-ReasoningAgent designed to evaluate LLM agents' reasoning, focusing on how they incorporate both fact and subjectivity when making decisions in cryptocurrency markets. This framework aims to clarify how LLMs reason through trading decisions, providing valuable insights that can guide future research in this field. We discuss the most related work here and leave more details in \autoref{supplementary-related} due to the limited space.


\vspace{-0.1cm}
\section{Conclusion}
\vspace{-0.1cm}
Our findings challenge the common assumption that stronger LLMs always outperform weaker ones, showing that advanced reasoning alone does not guarantee superior trading performance. To fully leverage the potential of stronger LLMs, we introduce FS-ReasoningAgent, a novel multi-agent framework that enhances decision-making by separating fact-based and subjectivity-based reasoning, thereby optimizing performance across various market conditions. Our experimental results demonstrate that FS-ReasoningAgent effectively harnesses the capabilities of stronger LLMs, achieving superior returns and higher Sharpe ratios in diverse market scenarios. Notably, we observe that subjectivity plays a more critical role in bull markets, whereas factual analysis is paramount in bear markets. This work encourages the research community to rethink strategies for maximizing the reasoning potential of LLMs, highlighting that without a carefully designed framework tailored to specific applications, advanced reasoning capabilities may remain underutilized.

\bibliography{iclr2025_conference}
\bibliographystyle{iclr2025_conference}

\clearpage
\appendix
\section*{Appendix}

\section{Experimental Environment} \label{env}
Our experiments were conducted using four NVIDIA H100 PCIe GPUs, managed by the NVIDIA-SMI 555.42.06 driver and leveraging CUDA 12.5 for optimal performance. The models in these experiments were implemented using PyTorch 2.0.0 in Python 3.12.5, ensuring compatibility and efficient execution on this powerful hardware setup.


\section{Supplementary Related Work} \label{supplementary-related}

\textbf{Time-Series Forecasting for Financial Markets} Time-series forecasting has been a pivotal research area in financial markets. Initial studies focused on predicting stock prices using approaches such as machine learning \citep{leung2021promises, patel2014stock}, reinforcement learning \citep{lee2001stock}, and conventional time-series models \citep{herwartz2017stock}. The Long Short-Term Memory (LSTM) model has emerged as a key method due to its ability to effectively manage sequential data \citep{sunny2020deep}. With the growing adoption of blockchain and cryptocurrencies, these methods have been adapted to predict crypto asset prices \citep{khedr2021cryptocurrency}. Researchers have considered both on-chain data, such as historical transactions and trading volumes \citep{ferdiansyah2019lstm}, and off-chain data, including social media sentiment and news analysis \citep{abraham2018cryptocurrency, pang2019cryptocurrency}. Integrating these diverse data sources has proven effective in capturing the volatile nature of cryptocurrency markets. Moreover, Transformer-based models have gained traction, with state-of-the-art models such as Informer \citep{zhou2021informer}, AutoFormer \citep{wu2021autoformer}, PatchTST \citep{nie2022time}, and TimesNet \citep{wu2022timesnet} setting new benchmarks in time-series forecasting.

\noindent \textbf{Self-Reflective Language Agents} The Self-Reflective framework introduces an innovative approach for enabling autonomous learning through iterative self-assessment and continuous refinement \citep{madaan2024self}. Complementary efforts focus on automating prompt refinement \citep{pryzant2023automatic, ye2024investigating} and generating feedback to enhance reasoning abilities \citep{paul2023refiner}. A notable advancement is the "Reflexion" framework proposed by \citep{shinn2024reflexion}, which enhances language agents by leveraging linguistic feedback stored in an episodic memory buffer, bypassing traditional weight update methods. These developments highlight the potential of LLMs to learn from past experiences and improve through self-reflection. 

\section{Evaluation Metrics} \label{metrics}

\textbf{(1) Return} measures the overall performance of the trading strategy, calculated as $\frac{w^{end}-w^{start}}{w^{start}}$, where $w^{start}$ and $w^{end}$ denote the initial and final net worth respectively.

\noindent \textbf{(2) Sharpe Ratio:} The Sharpe Ratio measures the risk-adjusted return and is calculated as $\frac{\bar{r}-r_f}{\sigma}$, where $\bar{r}$ represents the average daily return, $\sigma$ denotes the standard deviation of daily returns, and $r_f$ is the risk-free return. We set $r_f$ to 0, consistent with common practices in standard trading scenarios \citep{cheng2024sociodojo}.

\noindent \textbf{(3) Daily Return Mean} reflects the average daily performance of the trading strategy over the trading period.

\noindent \textbf{(4) Daily Return Std} represents the standard deviation of daily returns, indicating the volatility and risk associated with the strategy’s daily performance.

\section{Experiments Using Single LLMs} \label{reimplement}

\subsection{Dataset Splits}
We base our experiments testing single LLMs' trading performance on the dataset CryptoTrade provides which covers several months, detailed in \autoref{tab:original-period-stats}. This dataset captures the recent market performance of BTC, ETH, and SOL, highlighting challenges in identifying market trends and volatility. We divide the dataset into validation and test sets, using the validation set to fine-tune model hyperparameters and the test set to evaluate model performance. 

\begin{table*}[h!]
    \centering
\caption{Dataset splits. Prices are in US dollars. In each split, the transaction days include the start date and exclude the end date. We evaluate the total profit on the end date.}
\resizebox{0.7\linewidth}{!}{
\begin{tabular}{clccrrr}
\hline
Type                 & Split           & Start      & End        & Open     & Close    & Trend    \\ \hline
\multirow{3}{*}{BTC} & Validation      & 2023-01-19 & 2023-03-13 & 20977.48 & 20628.03 & \textcolor{mhold}{-1.67\%}  \\
                     & Test   Bearish  & 2023-04-12 & 2023-06-16 & 30462.48 & 25575.28 & \textcolor{mdec}{-15.61\%} \\
                     & Test   Bullish  & 2023-10-01 & 2023-12-01 & 26967.40 & 37718.01 & \textcolor{minc}{39.66\%}  \\ \hline
\multirow{3}{*}{ETH} & Validation    & 2023-01-13 & 2023-03-12 & 1417.13  & 1429.60  & \textcolor{mhold}{0.88\%}   \\
                     & Test   Bearish  & 2023-04-12 & 2023-06-16 & 1892.94  & 1664.98  & \textcolor{mdec}{-12.24\%} \\
                     & Test   Bullish  & 2023-10-01 & 2023-12-01 & 1671.00  & 2051.76  & \textcolor{minc}{22.59\%}  \\ \hline
\multirow{3}{*}{SOL} & Validation      & 2023-01-14 & 2023-03-12 & 18.29    & 18.24    & \textcolor{mhold}{-0.27\%}  \\
                     & Test   Bearish  & 2023-04-12 & 2023-06-16 & 23.02    & 14.76    & \textcolor{mdec}{-36.08\%} \\
                     & Test   Bullish  & 2023-10-01 & 2023-12-01 & 21.39    & 59.25    & \textcolor{minc}{176.72\%} \\ \hline
\end{tabular}
}
\label{tab:original-period-stats}
\end{table*}

\subsection{Data and Code Source}
We utilize the data and code available from CryptoTrade's public GitHub repository: \url{https://github.com/Xtra-Computing/CryptoTrade}.

\subsection{Experiment Results}
The experiment results shown in \autoref{tab:btc-updated} and \autoref{tab:sol-updated} indicate that stronger LLMs, such as o1-mini and GPT-4o, do not consistently outperform either traditional strategies or even simpler LLM models in terms of total returns and risk-adjusted performance. 

For instance, while GPT-4o performs reasonably well in Bull markets (28.47\% total return on BTC and 115.18\% on SOL), it fails to deliver the best results, trailing behind the simpler o1-mini model in BTC (36.50\%) and behind the traditional SLMA strategy on SOL (169.98\%). Furthermore, in Bear markets, o1-mini experiences significant reduction, with a -15.81\% return on BTC and -25.68\% on SOL, worse than the performance of weaker models like GPT-3.5-turbo. This pattern suggests that stronger LLMs, despite their advanced reasoning capabilities, do not necessarily make better trading decisions under all conditions, particularly in managing risk during downturns. Simpler models, such as GPT-3.5-turbo, and traditional strategies like SLMA, show better resilience and overall balanced performance across different market conditions, highlighting that more advanced LLMs may not always lead to superior results.

\begin{table}[h!]
    \centering
    \small
    \caption{Performance comparison of single LLMs, and baseline trading strategies on BTC during both Bull and Bear market conditions.}
    \resizebox{0.65\columnwidth}{!}{%
    \begin{tabular}{l|cc|cc|cc}
        \toprule
        \textbf{Strategy} & \multicolumn{2}{c|}{\textbf{Total Return (\%)}} & \multicolumn{2}{c|}{\textbf{Daily Return (\%)}} & \multicolumn{2}{c}{\textbf{Sharpe Ratio}} \\
        \cmidrule(lr){2-3} \cmidrule(lr){4-5} \cmidrule(lr){6-7}
        & \textbf{Bull} & \textbf{Bear} & \textbf{Bull} & \textbf{Bear} & \textbf{Bull} & \textbf{Bear} \\
        \midrule
        Buy and Hold & 39.66 & -15.61 & \ms{0.56}{2.23} & \ms{-0.24}{2.07} & 0.25 & -0.11 \\
        SMA & 22.58 & -21.74 & \ms{0.35}{1.89} & \ms{-0.36}{1.25} & 0.18 & -0.29 \\
        SLMA & 38.53 & -7.68 & \ms{0.55}{2.21} & \ms{-0.11}{1.23} & 0.25 & -0.09 \\
        MACD & 13.57 & -9.51 & \ms{0.22}{1.45} & \ms{-0.14}{1.56} & 0.15 & -0.09 \\
        Bollinger Bands & 2.97 & -1.17 & \ms{0.05}{0.32} & \ms{-0.02}{0.51} & 0.15 & -0.03 \\
        \midrule
        GPT-3.5-turbo & 18.84 & -9.12 & \ms{0.30}{1.69} & \ms{-0.14}{1.52} & 0.18 & -0.09 \\         
        GPT-4 & 26.35 & -11.72 & \ms{0.40}{1.76} & \ms{-0.18}{1.67} & 0.23 & -0.11 \\
        GPT-4o & 28.47 & -13.71 & \ms{0.43}{1.89} & \ms{-0.21}{1.71} & 0.23 & -0.12 \\ 
        o1-mini & 36.50 & -15.81 & \ms{0.53}{2.17} & \ms{-0.25}{1.94} & 0.25 & -0.13 \\ 
        \bottomrule
    \end{tabular}%
    }
    \label{tab:btc-updated}
\end{table}

\begin{table}[h!]
    \centering
    \small
    \caption{Performance comparison of single LLMs, and baseline trading strategies on SOL during both Bull and Bear market conditions.}
    \resizebox{0.65\columnwidth}{!}{%
    \begin{tabular}{l|cc|cc|cc}
        \toprule
        \textbf{Strategy} & \multicolumn{2}{c|}{\textbf{Total Return (\%)}} & \multicolumn{2}{c|}{\textbf{Daily Return (\%)}} & \multicolumn{2}{c}{\textbf{Sharpe Ratio}} \\
        \cmidrule(lr){2-3} \cmidrule(lr){4-5} \cmidrule(lr){6-7}
        & \textbf{Bull} & \textbf{Bear} & \textbf{Bull} & \textbf{Bear} & \textbf{Bull} & \textbf{Bear} \\
        \midrule
        Buy and Hold  & 176.72 & -36.08 & \ms{1.83}{6.00} & \ms{-0.61}{3.45} & 0.30 & -0.18 \\
        SMA           & 119.37 & 1.04  & \ms{1.43}{5.67} & \ms{0.02}{0.10} & 0.25 & 0.16 \\
        SLMA          & 169.98 & -8.11 & \ms{1.78}{5.93} & \ms{-0.11}{1.88} & 0.30 & -0.06 \\
        MACD          & 23.25  & -21.07 & \ms{0.35}{1.76} & \ms{-0.33}{2.44} & 0.20 & -0.13 \\
        Bollinger Bands & 2.92  & -21.69 & \ms{0.05}{0.35} & \ms{-0.35}{1.75} & 0.13 & -0.20 \\
        \midrule
        GPT-3.5-turbo & 102.45 & -24.08 & \ms{1.26}{4.54} & \ms{-0.39}{2.60} & 0.28 & -0.10 \\ 
        GPT-4 & 99.84 & -19.55 & \ms{1.24}{4.53} & \ms{-0.31}{2.35} & 0.27 & -0.13 \\
        GPT-4o & 115.18 & -16.32 & \ms{1.38}{4.98} & \ms{-0.25}{2.35} & 0.28 & -0.10 \\
        o1-mini & 102.67 & -25.68 & \ms{1.30}{5.27} & \ms{-0.41}{2.85} & 0.25 & -0.15 \\
        \bottomrule
    \end{tabular}%
    }
    \label{tab:sol-updated}
\end{table}




\section{Data Collection Details}
\label{collection.details}
The specific details of the data are as follows:

\begin{itemize}[leftmargin=*]
\item \noindent \textbf{Statistics:} We collect historical data from CoinMarketCap\footnote{\url{https://coinmarketcap.com}}, which provides daily insights into prices, trading volumes, and market capitalization of BTC, ETH, and SOL. For each day, we collect the opening price, closing price, transaction volume, average gas fees, the number of unique addresses, and the total value transferred on the cryptocurrency.

\item \noindent \textbf{News:} We employ the Gnews API\footnote{\url{https://pypi.org/project/gnews/}} to collect the news. The news dataset includes articles related to the cryptocurrencies, including BTC, ETH, and SOL, to ensure comprehensive and diverse coverage. The process begins by defining daily intervals within the specified date range. For each day, relevant English-language news articles are retrieved using cryptocurrency names as keywords, focusing on reputable sources like Bloomberg, Yahoo Finance, and crypto.news. This approach ensures a reliable and well-organized dataset for analyzing cryptocurrency news and market developments.
\end{itemize}

\section{Data Ethics} \label{collection.ethics}

\subsection{Statistical Data}
We obtain cryptocurrency statistical data from CoinMarketCap\footnote{\url{https://coinmarketcap.com}} and Dune\footnote{\url{https://dune.com/home}}. In line with CoinMarketCap's Terms of Service\footnote{\url{https://coinmarketcap.com/terms/}}, we are provided with a limited, personal, non-exclusive, non-sublicensable, and non-transferable license to access and use the content and services solely for personal purposes. We strictly refrain from using the service or its content for any commercial activities, complying fully with these terms. As for Dune’s Terms of Service\footnote{\url{https://dune.com/terms}}, we are allowed to access Dune’s APIs to perform SQL queries on blockchain data.

\subsection{News}
We utilize Gnews\footnote{\url{https://pypi.org/project/gnews/}} to systematically collect cryptocurrency-related news articles. In accordance with Gnews' Terms of Service\footnote{\url{https://gnews.io/terms/}}, we are allowed to download news for non-commercial, temporary viewing only. We are prohibited from modifying or copying the content, using it for commercial purposes or public displays, attempting to reverse engineer any software from Gnews, removing any copyright notices, transferring the content to others, or mirroring it on another server. We ensure that these conditions are strictly followed in our dataset. 







\section{Baselines} \label{baselines}
\begin{enumerate}
    \item \textbf{Buy and Hold:} A straightforward strategy where an asset is purchased at the beginning of the period and held until its end.
    
    \item \textbf{SMA \citep{gencay1996non}:} The Simple Moving Average (SMA) strategy makes buy and sell decisions by comparing the asset's price to its average over a specified period. We experiment with different time windows $[5, 10, 15, 20, 25, 30]$, selecting the period that performs best on a validation dataset.
    
    \item \textbf{SLMA \citep{wang2018predicting}:} The Staggered Moving Average (SLMA) method uses two moving averages with distinct durations. Trades are triggered when these averages cross. We evaluate various combinations of short and long moving averages, optimizing them based on validation set outcomes.
    
    \item \textbf{MACD \citep{wang2018predicting}:} The Moving Average Convergence Divergence (MACD) strategy identifies buy and sell signals by analyzing momentum shifts. It calculates the difference between a 12-day and a 26-day Exponential Moving Average (EMA), with a 9-day EMA acting as a trigger line. EMAs assign greater significance to recent data points.
    
    \item \textbf{Bollinger Bands \citep{day2023profitability}:} This approach generates signals by observing how the asset's price interacts with the Bollinger Bands, which consist of a 20-day SMA and bands placed at a set distance (typically two standard deviations) above and below. We adopt the standard settings for period length and band multiplier.

    \item \textbf{CryptoTrade \citep{li2024reflective}:} This strategy is an LLM-based trading agent designed specifically for cryptocurrency markets, expanding the typical application of LLMs beyond stock market trading. Experiments show that CryptoTrade outperforms time-series baselines in maximizing returns, though traditional trading signals still perform better under most of conditions.
\end{enumerate}

\section{Author Statement}
As authors of this paper, we hereby declare that we assume full responsibility for any liability or infringement of third-party rights that may come up from the use of our data. We confirm that we have obtained all necessary permissions and/or licenses needed to share this data with others for their own use. In doing so, we agree to indemnify and hold harmless any person or entity that may suffer damages resulting from our actions.


\section{Hosting Plan}
After careful consideration, we have chosen to host our code and data on GitHub. Our decision is based on various factors, including the platform's ease of use, cost-effectiveness, and scalability. We understand that accessibility is key when it comes to data management, which is why we will ensure that our data is easily accessible through a curated interface. We also recognize the importance of maintaining the platform's stability and functionality, and as such, we will provide the necessary maintenance to ensure that it remains up-to-date, bug-free, and running smoothly.

At the heart of our project is the belief in open access to data, and we are committed to making our data available to those who need it. As part of this commitment, we will be updating our GitHub repository regularly, so that users can rely on timely access to the most current information. We hope that by using GitHub as our hosting platform, we can provide a user-friendly and reliable solution for sharing our data with others.

\section*{Limitations}
One limitation of the FS-ReasoningAgent framework is its current focus on a limited number of cryptocurrencies, as it has been tested on individual assets. In the future, we plan to expand the framework to handle a diversified portfolio of cryptocurrencies, as well as explore its applicability to traditional financial markets, including stocks in the S\&P 500.

\section*{Broader Impacts}
Our research has several potential broader impacts beyond the scope of cryptocurrency trading. One important consideration is the risk that individuals might try to apply the trading strategies we discuss, leading to possible financial losses. We stress that the strategies presented are intended for academic research and experimental purposes only, and FS-ReasoningAgent is not designed or intended to offer investment advice.

Beyond the financial implications, our work encourages the broader research community to rethink the assumption that more powerful models always deliver better results in all contexts. By demonstrating that stronger LLMs may not outperform simpler models in certain tasks, we emphasize the need for careful model selection based on task-specific requirements.

\end{document}